\documentclass[journal]{IEEEtran}
\usepackage[T1]{fontenc}
\usepackage[latin9]{inputenc}
\usepackage{amsmath}
\usepackage{amssymb}
\usepackage{graphicx}
\usepackage[bookmarks=true,bookmarksnumbered=true,bookmarksopen=true,bookmarksopenlevel=1,
 breaklinks=false,pdfborder={0 0 1},backref=false,colorlinks=false]
 {hyperref}
\hypersetup{pdftitle={Your Title},
 pdfauthor={Your Name},
 pdfpagelayout=OneColumn, pdfnewwindow=true, pdfstartview=XYZ, plainpages=false}

\makeatletter
\let\oldforeign@language\foreign@language
\DeclareRobustCommand{\foreign@language}[1]{%
  \lowercase{\oldforeign@language{#1}}}

\usepackage[caption=false,font=footnotesize]{subfig}
\usepackage{braket}
\usepackage{cite}

\usepackage{tikz}
\usetikzlibrary{calc}

\usepackage{colortbl}

\makeatother

\begin{document}
\title{A Block Least Mean Square Method for Fiber Longitudinal Power Profile
Monitoring}
\author{Paolo Serena,~\IEEEmembership{Senior Member,~IEEE,} Chiara Lasagni,~\IEEEmembership{Member,~IEEE,}
Alberto Bononi, \IEEEmembership{Senior Member,~IEEE,} Fabien Boitier,~\IEEEmembership{Member,~IEEE},
and Joana Girard-Jollet\thanks{Manuscript received {*}{*}{*}{*}{*}{*}{*} {*}{*}, 2026. This work
has been supported by University of Parma through the action B\emph{ando
di Ateneo 2023 per la ricerca}. }\thanks{P. Serena, C. Lasagni, and A. Bononi are with the Department of Engineering
and Architecture, Universit\`a degli Studi di Parma, Parma, 43124,
Italy, and with the CNIT national laboratory of advanced optical fibers
for photonics (FIBERS) (e-mail: \protect\href{http://paolo.serena@unipr.it}{paolo.serena@unipr.it};
\protect\href{http://chiara.lasagni@unipr.it}{chiara.lasagni@unipr.it};
\protect\href{http://alberto.bononi@unipr.it}{alberto.bononi@unipr.it}).
F. Boitier and J. Girard-Jollet are with Nokia Bell Labs, Massy, 91300,
France (e-mail: \protect\href{http://fabien.boitier@nokia-bell-labs.com}{fabien.boitier@nokia-bell-labs.com};
\protect\href{http://joana.girard_jollet@nokia.com}{joana.girard\_jollet@nokia.com}).\\ This
work has been submitted to the IEEE for possible publication. Copyright
may be transferred without notice, after which this version may no
longer be accessible.}\thanks{Color versions of one or more of the figures in this paper are available
online at http://xxxxxxxxxx.xxxx.xxx.}\thanks{Digital Object Identifier xx.xxxx/JLT.xxxx.xxxxxxx}}
\markboth{Journal of Lightwave Technology}{Your Name \MakeLowercase{\emph{et al.}}: Your Title}
\IEEEpubid{}
\maketitle
\begin{abstract}
We propose a block least mean square (LMS) algorithm to monitor the
longitudinal power profile of a fiber-optic link through receiver-based
digital data from a coherent detector. Compared to the benchmark least
squares (LS) method, the proposed algorithm does not require large
matrix inversions or batch processing, thus allowing the received
data to be processed in blocks of minimum size by an overlap-save
algorithm, reducing complexity and latency. We propose an efficient
implementation of the method with a stochastic gradient update leveraging
a key computation in the frequency domain, offering computational
savings over state-of-the-art monitoring techniques. We test the proposal
in different scenarios by means of numerical simulations.
\end{abstract}

\begin{IEEEkeywords}
Least mean square (LMS), longitudinal power profile monitoring, overlap-save 
\end{IEEEkeywords}

\IEEEpeerreviewmaketitle{}

\section{Introduction}

\IEEEPARstart{D}{igital} longitudinal monitoring techniques have
received a lot of interest in recent years because they open up new
possibilities in autonomous network control through post-processing
inference rather than by distributed intrusive optical hardware devices
\cite{Wang}. The core objective is to leverage received data from
a digital receiver to perform system identification, thereby offering
the possibility to reduce the interaction with a control plane. This
attribute is especially important in modern optical systems that operate
in disaggregated network environments provided by several suppliers
and links comprising diverse optical fibers. 

Longitudinal power profile estimation (PPE) is one of the most important
monitoring techniques, because the primary cause of performance degradation
is typically due to anomalous losses that occur during field propagation
\cite{Tanimura20,Sasai_LS24}. Such kind of anomalies induce soft
failures, which result in a gradual degradation of quality of transmission
(QoT) that does not immediately cause a complete service outage. A
quick and reliable understanding of them is critical in order to implement
proactive countermeasures before a soft failure turns into a hard
failure.

After the introduction of coherent detection in optical communication
products, the use of coherent receiver's data for monitoring linear
dispersive impairments such as group velocity dispersion (GVD), polarization
mode dispersion (PMD), and polarization dependent loss (PDL) captured
the attention of the community, see e.g., \cite{Hauske}. Tanimura
et al. \cite{Tanimura10,Tanimura20} were among the first to exploit
the nonlinear interference (NLI) caused by the fiber Kerr effect for
tackling the PPE problem. They were able to perform PPE using a correlation
measurement (CM) between the coherent detector real system output
and a digital twin output. Such a study revealed the relevance of
describing the digital twin using a perturbative approximation \cite{VannucciRP}
with an NLI that is linearly related to the power profile. This research
prompted a thorough investigation and inspired innovative solutions
to its core challenges \cite{Gleb21,Sasai_SRS,May21,Sena22,Hui22,May22,Maysubsea,Sasai22,Takahashi_PDL,SerenaOFC23,Hahn_DGD,Zhou_ecoc25inv,Boitier25,Pilori25}.
A fundamental result has been provided by Sasai et al. \cite{Sasai_LS24},
who showed that the PPE problem can be efficiently addressed using
a least squares (LS) technique, a standard approach for linear system
identification. Contrary to the CM method, the LS method does not
need a calibration stage and thereby inherently eliminates any systematic
bias. Moreover, it improves the spatial resolution compared to CM
\cite{Sasai_limits}. 
\begin{figure*}[tbh]
\begin{centering}
\includegraphics[width=0.33\linewidth]{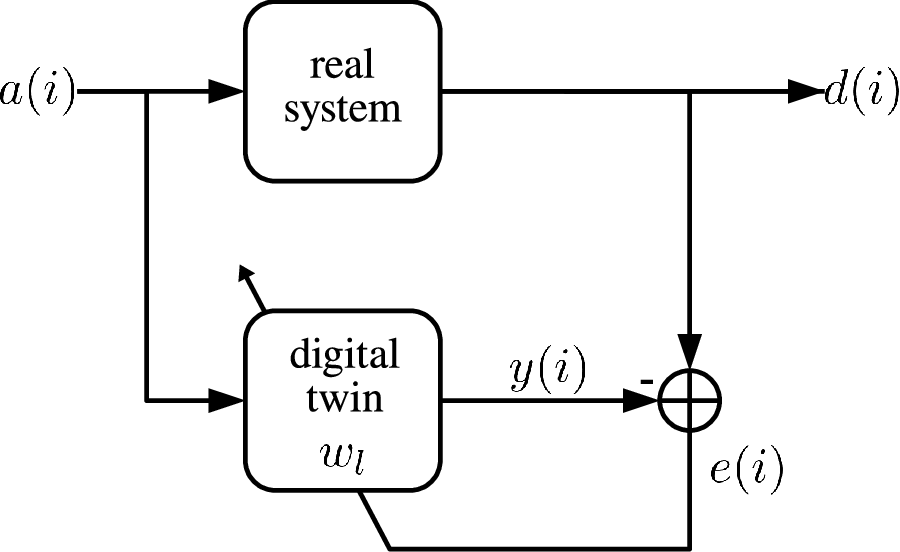}\includegraphics[width=0.33\linewidth]{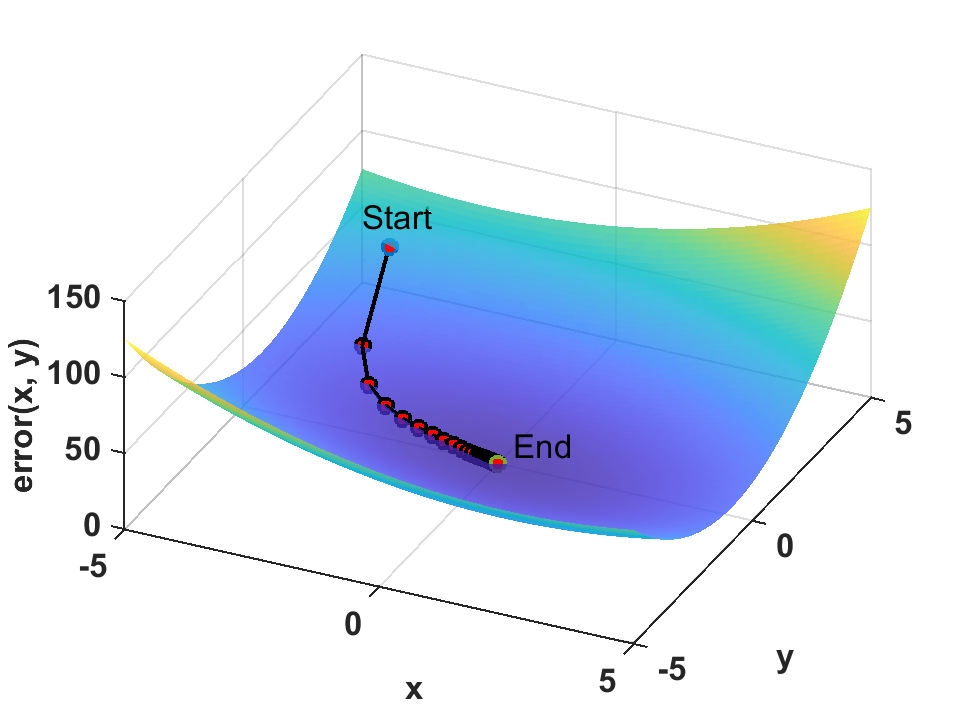}\includegraphics[width=0.33\linewidth]{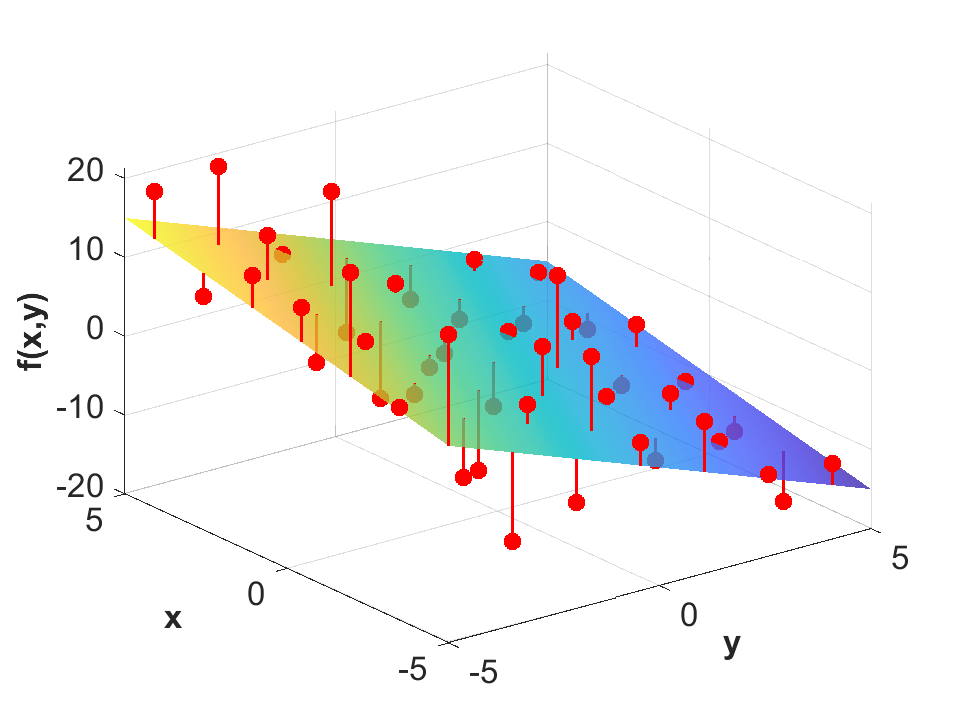}\\
\hspace{-2cm}\hfill(a)\hfill(b)\hfill(c)\hspace{2cm}
\par\end{centering}
\caption{(a): system identification problem setup. The objective is to find
the adaptive taps $w_{l}$ of the digital twin that minimize the error
$e$ between its output $y$ and the desired output $d$ in the presence
of noise. (b): LMS concept based on a stochastic gradient update using
the most recent observables. (c): Least squares method where all observables
(red dots) are used to fit the desired function, here a plane.\label{fig:sketch_LMSandLS}}
\end{figure*}

Although the least squares technique is one of the most efficient
in linear adaptive filter theory in terms of convergence speed, it
is complex since it requires the manipulation of a large matrix\textendash particularly
the computation of its inverse\textendash and must be implemented
within a resource-intensive digital twin architecture. Furthermore,
it uses batch processing to handle the full dataset without adaptivity
in its fundamental form. For these reasons, processing short blocks
of data is desirable in a real-world setting because it reduces processing
cost while successfully tracking a time-varying environment at reduced
latency. Solutions to overcome the least squares issues exist, such
as the recursive least squares (RLS) algorithm \cite{Haykin} or averaging
several batches of data. However, the first is even more complex,
while the second may introduce some artifacts \cite{Tang24_acc}.
The problem of complexity is particularly challenging and may prevent
the implementation of the PPE inside a transceiver digital signal
processor (DSP), with a peripheral field-programmable gate array (FPGA)
or even a power-hungry computer being the preferred solution \cite{Tanimura21}.
To mitigate the complexity of the digital twin, researchers have explored
the application of machine learning techniques (see \cite{Wang} and
the references therein). However, the effectiveness of these approaches
is often limited in operational networks, both because of the huge
dataset required to track a dynamic network evolution, and also because
of lack of measurements \cite{Musumeci25}, since soft failures, which
are essential for training the models, occur only rarely in highly
reliable optical systems. 

The least mean squares (LMS) approach is arguably the most widely-used
algorithm in adaptive filter theory \cite{Haykin}. The reason is
related to its inherent simplicity, since the LMS completely avoids
the large matrix and batch processing problems associated with the
standard LS technique. Since the PPE problem in a perturbative  framework
is linear, in this work we fill a gap in the literature by adapting
the LMS algorithm to the PPE scenario. The implementation is not straightforward,
because GVD may impose significant memory requirements inside the
digital twin, for which efficient solutions exist provided to work
with block processing rather than with the sample-by-sample processing
of the standard LMS. Consequently, we use a block LMS approach, which
provides a trade off between the basic LMS and the LS algorithms.
However, we cannot directly adopt the block LMS theory since we face
a mixed space-time problem that is not analyzed in standard LMS theory
\cite{Haykin}. Our technique computes the stochastic gradient by
using a key connection between the frequency and time domains, allowing
for efficient fast Fourier transforms (FFTs) while limiting block
processing to the minimum size dictated by the link's GVD. 

The remainder of the paper is organized as follows. Section \ref{sec:Problem-statement}
discusses the system identification problem and the differences between
the analog twin and the digital twin we use. Section \ref{sec:LMS-based-on}
describes the proposed block LMS algorithm. Section \ref{sec:Numerical-results}
shows the numerical results of the algorithm in different scenarios.
Finally, in Section \ref{sec:Conclusions} we draw our main conclusions.

\section{Problem statement\label{sec:Problem-statement}}

The power profile estimation is a system identification problem that
can be generally represented as shown in Fig. \ref{fig:sketch_LMSandLS}
(a). If we have a digital twin that accurately replicates the core
behavior of the real system as much as possible, we have discovered
the unknown power profile of the link under investigation. In this
work, we concentrate on replicating the propagation in the optical
fibers, reserving the analysis of other devices for future studies.

Sasai et al. \cite{Sasai_LS24} used a perturbative approximation
of the nonlinear link \cite{VannucciRP,Serena13} to make the identification
problem linear, and solved it with a LS technique which is widely
regarded as one of the most accurate methods for linear systems. In
this work we focus on the other popular algorithm, the LMS, and adapt
it to the peculiar properties of the NLI. A comparison between the
two methods is sketched in Fig. \ref{fig:sketch_LMSandLS} (b\textendash c).
The LMS aims at minimizing an error signal iteratively by a sequence
of steps using a stochastic gradient approximation calculated on the
most recent data, like searching for a valley in the fog. The LS method,
on the other hand, attempts to fit a parametric function to the real
system response by minimizing the mean square error between them.
Thus the key difference is that the LMS, in its basic form, operates
sample-by-sample, while the least squares operates via batch processing.
In this work we used a block-LMS technique to slightly relax the sample-by-sample
feature in favor of fast convolutions through FFT. Before introducing
it, we discuss the relation between a digital twin and a standard
analog twin.

\subsection{Analog twin}

Given the transmitted electric field $A(t)$, a reasonable approximation
of the received field using a first-order regular perturbation (RP)
approximation is:
\begin{equation}
A_{\text{out}}(t)\approx A(t)+n_{\text{SPM}}(t)+r(t)\label{eq:rp}
\end{equation}
with $n_{\text{SPM}}$ the NLI caused by self-phase modulation (SPM)
and $r(t)$ the remaining noise, such as amplified spontaneous emission
(ASE) and cross-channel NLI. According to the RP method \cite{VannucciRP},
the NLI is the result of a linear accumulation along the distance:
\begin{equation}
n_{\text{SPM}}(t)=-j\int\gamma^{\prime}(z)u(z,t)\text{d}z\label{eq:NLIanalog}
\end{equation}
where, as in \cite{Sasai_LS24}, we introduced $\gamma^{\prime}(z)\triangleq\gamma(z)f(z)$,
with $\gamma(z)$ the fiber nonlinear coefficient, $f(z)$ the longitudinal
optical power profile, and 
\begin{equation}
u(z,t)\triangleq\left(g_{z}\otimes{\cal N}\left(h_{z}\otimes A\right)\right)(t)\,.\label{eq:u}
\end{equation}
In this formulation, the symbol $\otimes$ indicates temporal convolution,
whereas ${\cal N}(A)\triangleq||A||^{2}A$ is the nonlinear Kerr operator
with $||.||$ indicating the Euclidean norm. The symbol $h_{z}(t)$
is the impulse response of GVD from system input to link coordinate
$z$, while $g_{z}(t)$ is the impulse response of GVD from $z$ to
system end. 

Even at typical powers, the approximation (\ref{eq:rp}) could be
inadequate because the nonlinear Kerr effect manifests itself not
only as an additive term but also as a phase that may be far from
zero radians, thus weakening the Taylor approximation at the core
of the RP method. The enhanced RP (eRP) \cite{VannucciRP}, which
makes the transformation
\[
A_{\text{out}}(t)=e^{j\varphi}E(t)
\]
offers a better approximation by perturbing $E$ only:
\[
A_{\text{out}}(t)\approx e^{j\varphi}\left(A(t)+n_{\text{SPM}}(t)-j\varphi A(t)\right)+r(t)\,.
\]
While the term $\exp(i\varphi)$ is irrelevant to our discussion because
it is eliminated by the carrier phase recovery, the ``eRP correction''
given by $j\varphi A$ is crucial \cite{Sasai_limits}. 

A reasonable choice for $\varphi$ is the time average phase rotation
induced by the Kerr effect \cite{Serena13,SerenaTIWDC}. However,
since the eRP correction is linear in $\varphi$, we found it more
convenient to allow the LMS finding its best approximation for $\varphi$.
This idea generalizes the proposal of \cite{KimCPE23} within the
scope of the RP approach, and decreases the algorithm's dependency
on prior information.

\subsection{Digital twin\label{subsec:A-discrete-time-digital}}

At the optical level, the RP field (\ref{eq:u}) consists of ``linear
+ nonlinear + linear'' steps, with a GVD operator modeling each linear
step separately. However, because our algorithm must work at the digital
level, we actually need an RP approximation relating the transmitted/detected
digital symbols rather than the input/output analog field. The fully
digital twin starts with the observation that the transmitted signal
is usually linearly modulated with a symbol period $T$ by $A(t)=\sum_{i}a(i)p(t-iT)$,
which in Fourier domain is:
\begin{equation}
\tilde{A}(\omega)=\tilde{a}\left(e^{j\omega T}\right)\tilde{P}(\omega)\label{eq:DTFT}
\end{equation}
with $\tilde{a}\left(e^{j\omega T}\right)$ the discrete-time Fourier
transform (DTFT) of the discrete symbol sequence $a(i)$, whereas
$\tilde{A}(\omega)$ and $\tilde{P}(\omega)$ represent the Fourier
transforms of $A(t)$ and $p(t)$, respectively. In this framework,
$\tilde{P}(\omega)$ is an additional linear operation preceding the
mentioned steps. Similarly, matched filtering and GVD equalization
are extra linear operations that occur after those steps. Notably,
the matched filter has frequency response that is the conjugate of
the pulse's spectrum, $\tilde{P}^{*}(\omega)$, and the GVD equalization
similarly acts as a filter with conjugated frequency response of the
GVD response. This observation enables a relationship between the
impulse responses of $g_{z}$ and $h_{z}$ once including pulse shaping
and detection in the aforementioned steps \cite{mecozzi12}:
\begin{equation}
\tilde{g}_{z}\left(\omega\right)=\tilde{h}_{z}^{*}\left(\omega\right)=\tilde{P}(\omega)e^{-j\frac{\omega^{2}}{2}\int_{0}^{z}\beta_{2}(\xi)\text{d}\xi}\label{eq:ghconh}
\end{equation}
with $\beta_{2}$ the local dispersion coefficient of the fiber. Since
$g_{z}$ is sufficiently band- and time-limited, we can obtain a practical
representation in the frequency domain by sampling $p(t)$ in time
with $N_{\text{t}}$ samples per symbol, and work with its discrete
Fourier transform (DFT), while upsampling $a(i)$ to have a signal
over the same temporal grid before taking its DFT. Equation (\ref{eq:DTFT})
thus discretizes into:
\[
\text{DFT}\left[A\left(\frac{mT}{N_{\text{t}}}\right)\right]=\text{DFT}\left[a_{\uparrow N_{\text{t}}}(m)\right]\times\text{DFT}\left[p\left(\frac{mT}{N_{\text{t}}}\right)\right]
\]
with $m$ the sample index and $\uparrow N_{\text{t}}$ indicating
up-sampling by a factor $N_{\text{t}}$. We also need to discretize
the spatial coordinate $z$ in (\ref{eq:NLIanalog}). This can be
accomplished using quadrature formulas for numerical integration:
\begin{equation}
n_{\text{SPM}}(t)\approx\sum_{l}\rho_{l}\gamma^{\prime}(z_{l})u(z_{l},t)\label{eq:nliquad}
\end{equation}
where $z_{l}$ are the coordinates reproduced by the digital twin.
Typically, the grid is uniform, for which a good quadrature rule is
the mid-point rule\footnote{We recall that the mid-point rule is more accurate than the more common
trapezoidal rule without adding complexity.} with $\rho_{l}$ independent of $l$, i.e., the constant step-size.

In this framework, the digital twin builds the following signal at
symbol index $i$:
\begin{equation}
y(i)=\hat{a}(i)(1-j\varphi)+n(i)\label{eq:output_digtwin}
\end{equation}
where we identify $\hat{a}(i)$ as the detected symbols after the
receiver DSP, $\varphi$ as the eRP correction, and the NLI due to
SPM:
\begin{equation}
n(i)\triangleq\sum_{l=0}^{M-1}\rho_{l}\gamma^{\prime}(z_{l})\left(g_{l}\otimes{\cal N}\left(h_{l}\otimes A\right)\right)(iT)\label{eq:NLI}
\end{equation}
 where $M$ is the number of integration points and we introduced
$h_{l}\equiv h_{z_{l}}(t)$, $g_{l}\equiv g_{z_{l}}(t)$ to simplify
the notation. In the next section, we propose an efficient implementation
that exploits the overlap-save algorithm within the context of our
block LMS solution.

\section{Block LMS based on overlap-save\label{sec:LMS-based-on}}

System identification is generally accomplished by minimizing an error
function. Since the problem is stochastic, ensemble averaging of the
error function is commonly utilized in analytical works because it
is simple to manipulate in linear systems, as demonstrated by Wiener
in his seminal works \cite{Haykin}. Unfortunately, using ensemble
averaging is impractical in many problems, including the one we seek
to solve, because it requires the observation of many independent
realizations of the system. The common solution is to replace ensemble
averaging with deterministic averaging, which results in the least
squares method. LMS reduces deterministic averaging to its most basic
form, namely taking the current realization of the error. This simple
approximation has significant implications because it reduces the
complexity and allows for a recursive search for the minimum by a
stochastic gradient method, thus by recursively updating the result
by taking short steps in the steepest descent direction, see Fig.
\ref{fig:sketch_LMSandLS}(b). 

According to the previous Section, the input/output relation of the
optical link is linear in the longitudinal power profile and more
generally in the $\gamma^{\prime}(z_{l})$ coefficients. We collect
such coefficients into a vector of length $M$
\[
\mathbf{w}(i)=\left[\gamma^{\prime}(z_{0}),\gamma^{\prime}(z_{1}),\ldots,\gamma^{\prime}(z_{M-1})\right]^{T}
\]
and introduce a temporal dependence on the symbol $i$. The steepest
descent updating rule is \cite{Haykin}:
\[
\mathbf{w}(i+1)=\mathbf{w}(i)-\frac{\mu}{2}\nabla J(i)
\]
where $J$ is the error function that we want to minimize and the
gradient $\nabla$ is with respect to the taps $\mathbf{w}$. According
to Fig \ref{fig:sketch_LMSandLS}(a), the error is $e(i)=d(i)-y(i)$,
with $d(i)$ the desired response at time $i$ (the real system output)
and $y(i)$ the corresponding digital twin output. The steepest descent
method uses $J=\mathbb{E}\left[|e(i)|^{2}\right]$ while the LMS simply
$J=|e(i)|^{2}$. Unfortunately, the basic LMS rule works on a sample-by-sample
basis, preventing the use of efficient techniques like the FFT that
operate over blocks of data. However, the use of FFT is particularly
desirable because the NLI involves convolutions with filters that
may have a long duration. A block-LMS appears to be the most logical
solution to this problem. The idea is to update the taps not at each
sample but at each block of samples (mini batch), chosen longer than
the system memory, so that FFTs can be utilized. Nevertheless, the
error function $J$ varies across the block, hence it has been recommended
to use averaging to get a unique value \cite[chap. 8]{Haykin}. Using
$k$ to indicate block index, we thus use the following updating rule:
\begin{equation}
\mathbf{w}(k+1)=\mathbf{w}(k)-\frac{\mu}{2}\sum_{i=0}^{L-1}\nabla J(kL+i)\label{eq:update_block}
\end{equation}
where each block contains $L$ symbols. A significant distinction
exists between our proposed block LMS and the conventional block LMS
\cite{Haykin}: the former addresses a mixed space-time problem that
is time-invariant but space-variant with respect to the key variable
under investigation. Our goal is to leverage the time-invariant property
for efficient operations by expressing both the digital twin output
$y$ and the tap update rule with optimized computational methods. 

We begin our analysis with $y$. Thanks to the overlap-save algorithm
(see (\ref{eq:key_os})), we have the following relation between the
analog signal (\ref{eq:u}) and its discrete counterpart for $i=0,1,\ldots,L-1$:
\begin{multline*}
\left(g_{l}\otimes{\cal N}\left(h_{l}\otimes A\right)\right)((kL+i)T)=\\
\left(g_{l,\text{pad}}\circledast{\cal N}\left(h_{l,\text{pad}}\circledast\hat{A}_{k}\right)\right)_{\text{val}}(iN_{\text{t}})
\end{multline*}
where we converted an analog problem based on linear convolutions
$\otimes$ into a discrete one using efficient circular convolutions
$\circledast$. While we refer the reader to Appendix \ref{sec:Overlap-save-Algorithm}
for more details, $\hat{A}_{k}$ is a vector resulting from the concatenation
of the symbols in block $(k-1)$ and block $k$, subsequently upsampled
by a factor $N_{\text{t}}$. ``val'' refers the not-aliased part
of the vector, while ``pad'' means zero-padded. All vectors are
of length $(2L\times N_{\text{t}})$, where the factor two is consistent
with the block concatenation. Additionally, we observe that we are
employing multirate DSP, with increased sampling frequency to handle
optical fields and downsampling by a factor of $N_{\text{t}}$ for
the final output. 

To simplify the notation, we introduce a vector containing aliased
and non-aliased samples:
\[
U_{lk}\triangleq\left(g_{l,\text{pad}}\circledast{\cal N}\left(h_{l,\text{pad}}\circledast\hat{A}_{k}\right)\right)(n),\quad n=0,\ldots,2LN_{\text{t}}-1
\]
and exploit the properties of the DFT to transform (\ref{eq:NLI})
into:
\begin{multline*}
n(i)=\left(\text{DFT}^{-1}\Bigg[\sum_{l=0}^{M-1}\rho_{l}w_{l}(k)\cdot\text{DFT}\left[U_{lk}\right]\Bigg]\right)_{\text{val}}(iN_{\text{t}}),\\
i=0,\ldots,L-1
\end{multline*}
where $w_{l}$ is the $l$th element of $\mathbf{w}$. Note that we
return back to the time domain only after having computed the summation
in $l$, thus with only one final IFFT. Finally, the computation of
$y(m)$ follows straightforwardly from (\ref{eq:output_digtwin}).

Regarding the tap-update rule, the LMS transforms (\ref{eq:update_block})
into \cite{Haykin}:
\begin{equation}
\mathbf{w}(k+1)=\mathbf{w}(k)+\mu\text{Re}\left[\mathbf{v}(k)\right]\label{eq:update_block-1}
\end{equation}
where $\mathbf{v}$ is an $M\times1$ vector whose $j$th element
$v_{j}(k)$ is:
\begin{align*}
v_{j}(k) & =\sum_{i=0}^{L-1}e_{k}^{*}(i)\left(g_{j}\otimes{\cal N}\left(h_{j}\otimes A_{k}\right)\right)(iT)\\
 & =\sum_{i=0}^{L-1}e_{k}^{*}(i)\left(U_{jk}\right)_{\text{val}}(iN_{\text{t}})\\
 & =\sum_{n=0}^{2LN_{\text{t}}-1}e_{k,\text{int}}^{*}(n)U_{jk}(n)\\
 & =\frac{1}{2LN_{\text{t}}}\sum_{f=0}^{2LN_{\text{t}}-1}\tilde{e}_{k,\text{int}}^{*}(f)\tilde{U}_{jk}(f),\quad j=0,1,\ldots,M-1
\end{align*}
where $e_{k}(i)\triangleq e(kL+i)$ is the error in block $k$, tilde
indicates DFT, and we used the Parseval's theorem \cite{Manolakis}
in the last identity, which is crucial for an efficient implementation.
$e_{k,\text{int}}$ is a vector resulting from upsampling $e_{k}(i)$
by a factor $N_{\text{t}}$ and subsequently applying zero-padding
specifically at the indexes where $U_{jk}$ contains non-valid samples,
allowing us to process the full $U_{jk}$ vector rather than the subset
$(U_{jk})_{\text{val}}$ and thus the use of Parseval's theorem. 
This approach allows the computation of $g_{j}\otimes{\cal N}\left(h_{j}\otimes A_{k}\right)$
in the frequency domain in the local $k$th window, thereby saving
$M$ IFFT at the cost of one extra FFT for $e_{k,\text{int}}$.

We found it useful to update the eRP correction $\varphi$ independently
of the taps $\mathbf{w}$, with a new control on its step size, called
$\mu_{0}$. Hence we treat it as an independent finite impulse response
(FIR) filter with one tap, $\varphi$, whose update rule follows similarly:
\begin{equation}
\varphi(k+1)=\varphi(k)+\mu_{0}\text{Im}\left[\sum_{i=0}^{L-1}e_{k}^{*}(i)\hat{a}(kL+i)\right]\label{eq:phi_erp}
\end{equation}
where the imaginary part is strictly related to the $j$ in front
of $\varphi$ in (\ref{eq:output_digtwin}).

\subsection{Block diagram}

\begin{figure*}[tbh]
\begin{centering}
\includegraphics[width=0.8\linewidth]{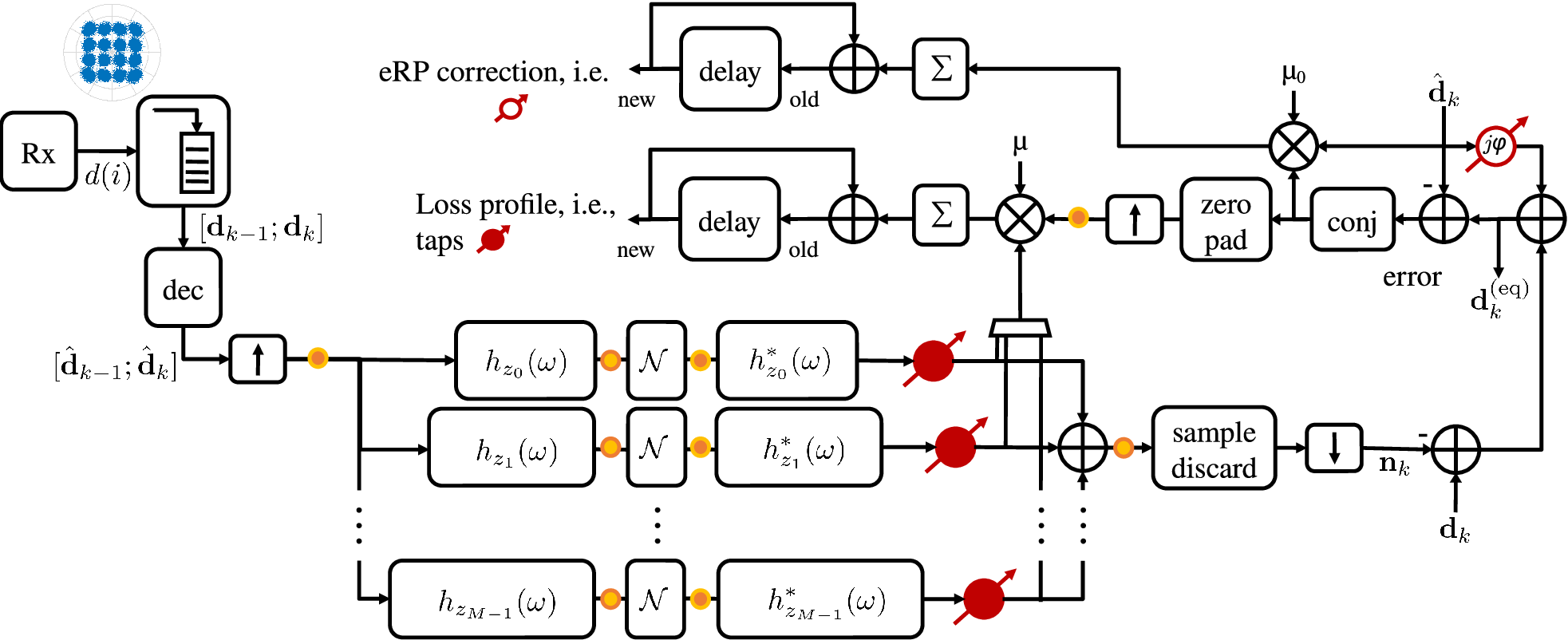}
\par\end{centering}
\caption{LMS block diagram for monitoring the longitudinal power profile (represented
by the filled taps) at coordinates $z_{l}$. The dark circle with
a light ring represents FFT, while inverted colors indicate IFFT.
An equalized replica of the data, $\mathbf{d}_{k}^{(\text{eq})}$,
is a concomitant outcome of the monitoring procedure. \label{fig:LMS-block-diagram}}

\end{figure*}

A block diagram of the proposed LMS power profile monitor is shown
in Fig. \ref{fig:LMS-block-diagram}. The Rx block at the top-left
corner outputs the received noisy symbols from a typical receiver-DSP
including matched filtering, equalization, time recovery, frequency
and carrier phase recovery. The Rx symbols are then buffered into
vectors 
\[
\mathbf{d}_{k}\triangleq[d(kL),\ldots,d(kL+L)]^{T}
\]
describing the blocks. These blocks are then concatenated in pairs,
i.e., $[\mathbf{d}_{k-1};\mathbf{d}_{k}]$, as per the overlap-save
idea. The block length $L$ must exceed the duration of the system
memory, which is set by GVD and the matched filter duration. As a
rule of thumb, the memory that GVD adds in a fiber of length $z$
is $2\pi|\beta_{2}|z/T^{2}$ symbols. 

The module ``dec'' corresponds to decision and potentially includes
forward-error correction, provided that its challenges, in particular
latency, can be addressed \cite{Chang25,Andrenacci25,Jiang25}. Then,
the signal is upsampled to create the vector $\hat{A}_{k}$. We used
upsampling of $N_{\text{t}}=2$ samples per symbol, which is enough
to capture the pulse bandwidth. It is worth noting that aliasing occurs
within the nonlinear block when the pulse roll-off is greater than
zero. However, our results with roll-off of $0.1$ did not show any
significant aliasing. 

After upsampling we use FFT to move into the frequency domain, which
is shown as a dark orange circle with a light yellow ring for notation
purposes (and an inverted scheme for IFFT). The corresponding DFT
is then propagated into the digital twin through a sequence of parallel
``linear+nonlinear+linear'' steps at the heart of the RP approximation,
as explained in Section \ref{subsec:A-discrete-time-digital}.  Each
local contribution is then weighted by the the corresponding tap $w_{l}$,
which is the primary objective of our algorithm. Note that we return
back to the time domain by an IFFT only after summing all the NLI
contributions in the frequency domain, thus significantly saving on
FFT operations. This is possible because we calculate the error function
that drives the LMS in the frequency domain. Similarly, we performed
a single FFT of $\hat{A}_{k}$ at the beginning, thus limiting the
computational effort to one FFT and one IFFT per path.

After returning to the time domain, the aliased samples from the overlap-save
algorithm are discarded (operation indicated by ``val'' in our analysis),
and the signal is then downsampled back to one sample per symbol.
We thus have an estimation of the NLI (\ref{eq:NLI}) within the $k$th
block of symbols, indicated by $\mathbf{n}_{k}$. We then remove the
SPM-NLI and the eRP correction from the received noisy symbols to
obtain the cleaner signal $\mathbf{d}_{k}^{(\text{eq})}$, as we have
effectively created a spatially resolved Volterra nonlinear equalizer
that functions alongside the monitoring device. Finally, we calculate
the error $e(i)=d(i)-y(i)$ for each index $i$ in the block and update
the taps of the eRP correction and of the $\gamma^{\prime}$ coefficients
block-by-block as described in the previous Section. Note that the
last operation exploits a frequency domain computation after zero-padding
and summation thanks to Parseval's theorem. 

Even though the block diagram refers to a single signal, with dual
polarization we just duplicate the operations. However, if PMD and
PDL  are small, we recommend leveraging the correlated NLI in the
two polarizations by averaging their error functions to mitigate the
noise.

\subsection{Polyphase implementation}

\begin{figure}[tbh]
\begin{centering}
\includegraphics[width=0.8\linewidth]{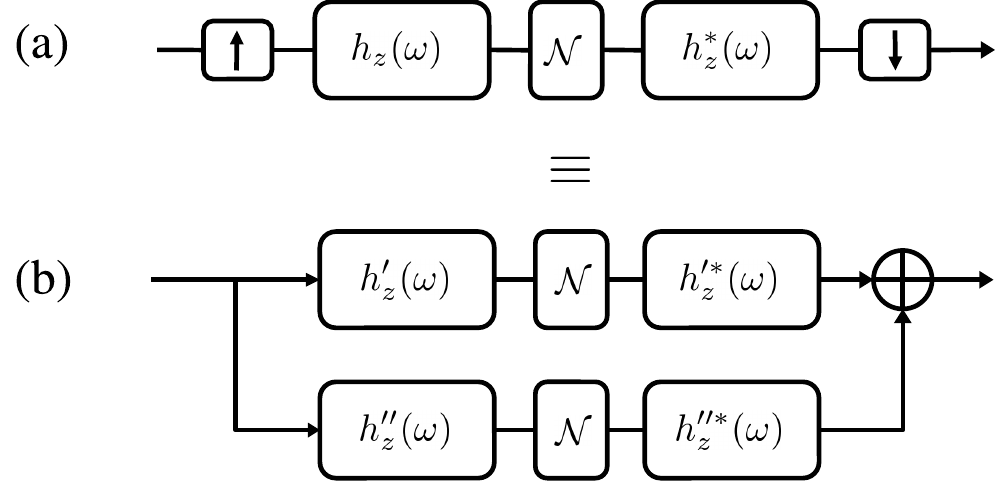}
\par\end{centering}
\caption{(a) One branch of the block diagram shown in Fig. \ref{fig:LMS-block-diagram}.
(b) Polyphase implementation of the branch. The two structures are
equivalent with similar complexity; however the polyphase one works
at a halved sampling rate.\label{fig:polyphase}}

\end{figure}

The upsampler and downsampler can be removed when using a polyphase
implementation with additional parallelization \cite{Manolakis}.
Figure \ref{fig:polyphase} illustrates the equivalence between a
single branch of the digital twin and its polyphase structure using
$N_{\text{t}}=2$. The impulse responses $h_{z}^{\prime}(i)$ and
$h_{z}^{\prime\prime}(i)$ contain the odd and even samples of $h_{z}(i)$,
respectively. The main advantage of a polyphase representation is
a reduced sampling rate (one sample per symbol) achieved by using
smaller, parallel filters instead of one large filter. When using
the polyphase structure, we can also remove the extra upsampler of
the error signal in Fig. \ref{fig:LMS-block-diagram}, thus applying
Parseval's theorem on the DFT of a signal sampled at one sample per
symbol.

\section{Numerical results\label{sec:Numerical-results}}

We tested the proposed LMS algorithm on a dual-polarization 16 quadrature
amplitude modulated (QAM) channel at $64$ Gbd. The ``real system''
was modeled by a numerical simulation with the split-step Fourier
method (SSFM), by launching sequences of $65536$ symbols shaped with
root-raised-cosine pulses of roll-off $0.1$ into an optical link
with $100$\textendash km spans and lumped amplification. The SSFM
step was updated with the constant local error (CLE) criterion \cite{Musetti}
with a first step of $200$ m. The optical fibers had nominal loss
coefficient of $0.2$ dB/km, dispersion $17$ $\text{ps}\cdot\text{nm}^{-1}\text{km}^{-1}$,
nonlinear coefficient $\gamma=1.26$ W$^{-1}$km$^{-1}$. PMD and
PDL were absent. We assumed the LMS to have perfect knowledge of the
fiber dispersion and of the Tx symbols (except one result). To test
the monitoring capabilities, we inserted one or multiple lumped loss
anomalies at different positions of the link depending on the setup.
Since ASE and cross-channel Kerr effects act as noise for the algorithm,
we simulated them through receiver noise loading of an additive white
Gaussian noise (AWGN) source, $r(t)$ in our analysis, with the signal-to-noise
ratio (SNR) as a control variable. We used a basic matched-filter
receiver with GVD and average carrier phase recovery.

We started by testing the algorithm in the absence of noise to test
its capabilities. We analyzed a multi-anomaly case in a single span,
including: i) six loss anomalies of $0.25$ dB each, and ii) backward
Raman amplification. Signal power was $P=5$ dBm, while we iterated
the LMS over $1000$ independent realizations of the real system.
In (\ref{eq:update_block}) we used $\mu=\bar{\mu}/P^{5/2}$ with
$\bar{\mu}=0.05$, where the normalization to $P^{5/2}$ is an attempt
to establish a general rule of thumb. The scaling was motivated by
a predicted scaling with $P^{3}$ of the NLI variance \cite{Serena13}
and a scaling with $P^{2}$ of the error variance on the (normalized)
constellation. On the other hand, in (\ref{eq:phi_erp}) we used $\mu_{0}=2\cdot10^{-4}\Phi$
with $\Phi$ the average nonlinear phase accumulated in the link.
The spatial resolution of the digital twin was $5$ km.

\begin{figure}[tbh]
\begin{centering}
\includegraphics[width=0.8\linewidth]{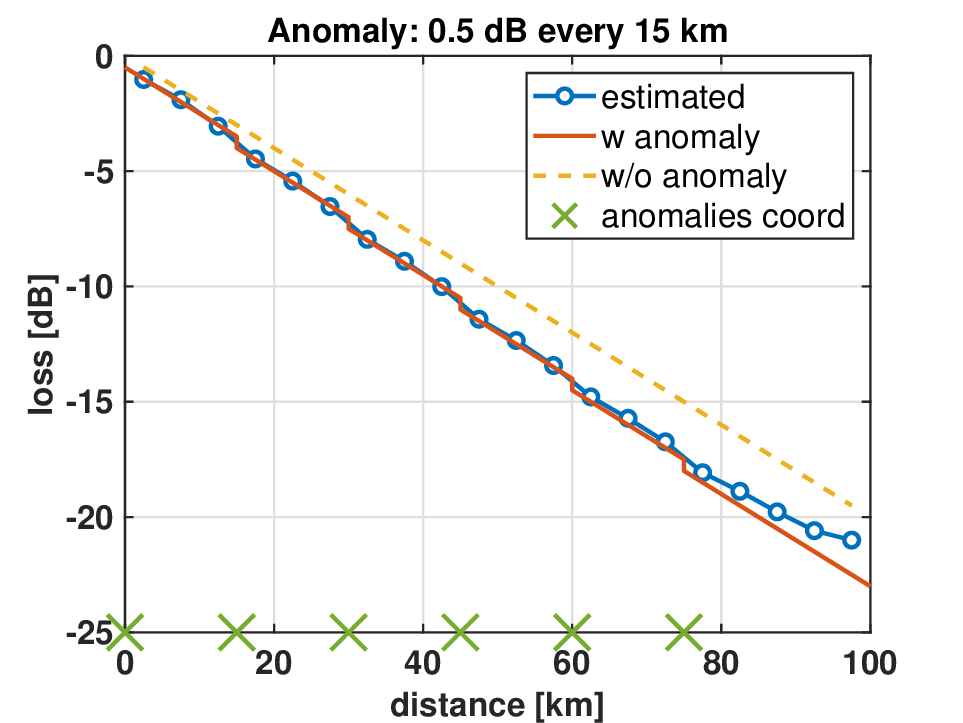}
\par\end{centering}
\begin{centering}
\includegraphics[width=0.8\linewidth]{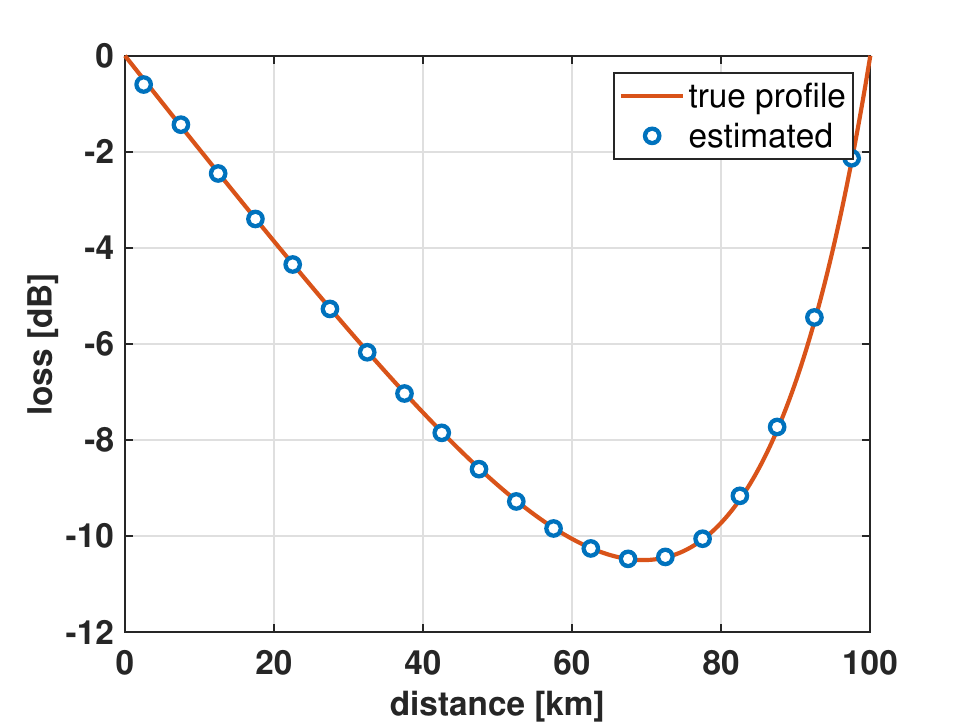}
\par\end{centering}
\caption{Top: A multianomaly case with six loss anomalies of $0.25$ dB every
$15$ km. Bottom: net loss profile in the presence of backward Raman
pumping. No additive noise.\label{fig:Multianom_and_Raman}}

\end{figure}

Figure \ref{fig:Multianom_and_Raman} shows the true loss profile
and the predicted one by the LMS algorithm. The match is excellent,
with a slight overestimation in the low power regime, a problem common
to any NLI-based technique where the useful signal is soaked in noise.

Like any LMS method, the normalized step size $\bar{\mu}$ is a trade
off between speed of convergence and accuracy. To test such trade-off,
we analyzed a three-span link with anomaly of $1$ dB after $125$
km, hence roughly after one effective length in the second span. In
this simulation we employed an exceptionally low SNR of 10 dB to stress
the behavior under an extreme situation. For reference, a 16-QAM signal
in an AWGN channel at SNR=10 dB has at best a mutual information of
3.27 bits/symbol, smaller than its 4 bits/symbol nominal value, thus
requiring a forward error correction (FEC) redundancy of at least
0.73 bits/symbol. As a quality metric, we estimated the root mean
square error (RMSE) between the estimated profile and the true one
in dB scale. As observed previously, since perturbative methods fail
in the low power regions, we decided to reject results at coordinates
with a path loss greater than $15$ dB. This ensures that excess errors
do not undermine the meaning of the RMSE.

\begin{figure}[tbh]
\begin{centering}
\includegraphics[width=0.8\linewidth]{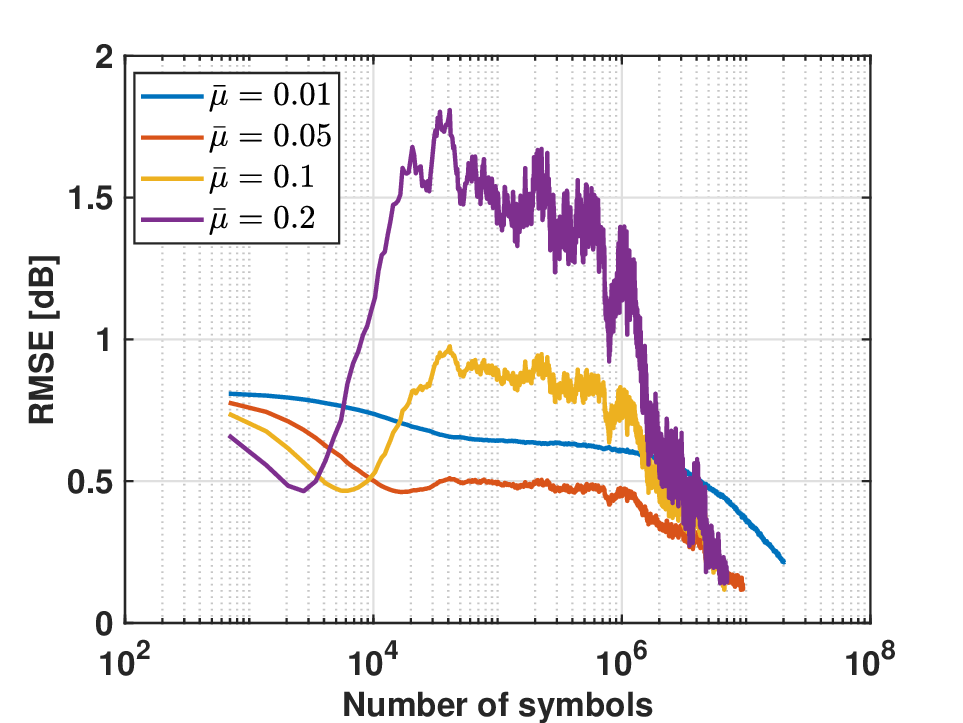}
\par\end{centering}
\caption{RMSE vs. the number of symbols processed by the LMS algorithm at different
normalized LMS step sizes $\bar{\mu}$. Three-span link with anomaly
of $1$ dB after $125$ km. Taps initialized with the nominal fiber
loss. \label{fig:RRMSE_vs_nubsampandmu}}

\end{figure}

The RMSE is shown in Fig. \ref{fig:RRMSE_vs_nubsampandmu}. To provide
context for the RMSE, keep in mind that knowledge of only the nominal
loss profile excluding anomalies yields an RMSE in this setup of $0.5$
dB. We observe an improvement over this benchmark, with the RMSE
eventually decreasing as the number of samples increases. The estimation
becomes unstable at $\bar{\mu}=0.2$, even though it is able to converge
reasonably well. It is worth mentioning that while any starting value
for the filter taps is feasible, different initial setups will result
in different convergence rates and paths. In our simulations, we initialized
the taps with knowledge of the nominal fiber loss, which allowed for
faster convergence than a random or zero-initialized start. 

\begin{figure}[tbh]

\begin{centering}
\includegraphics[width=0.8\linewidth]{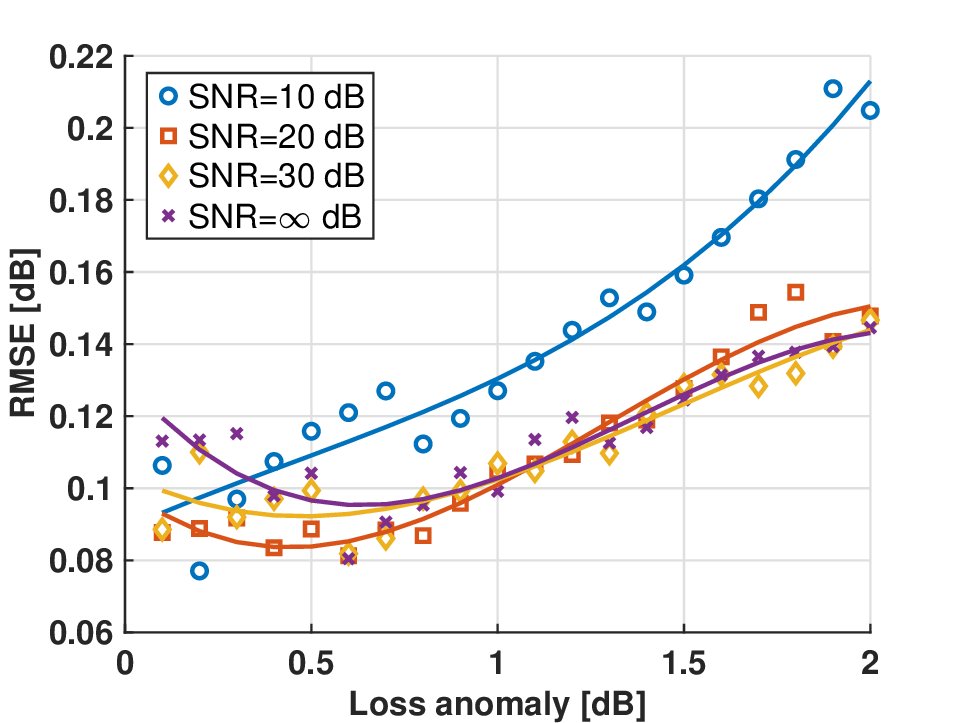}
\par\end{centering}
\begin{centering}
\includegraphics[width=0.8\linewidth]{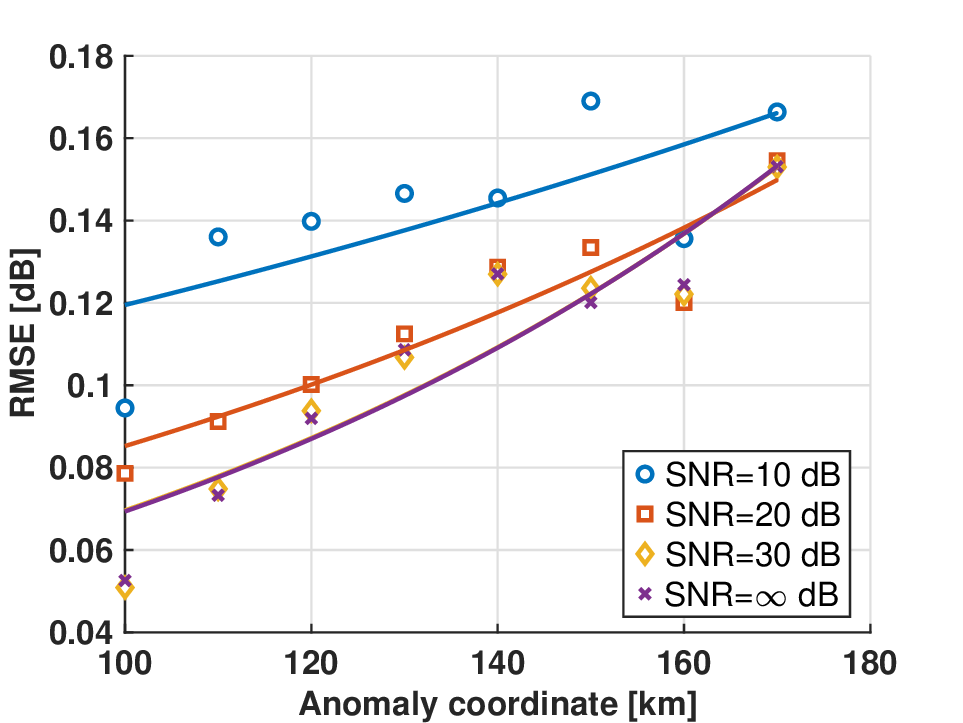}
\par\end{centering}
\caption{RMSE in the three-span system. Top: single loss anomaly of variable
magnitude located at 125 km. Bottom: loss anomaly of 1 dB at a variable
location. The lines are interpolations to guide the eye. \label{fig:var_lossan_zan}}
\end{figure}

To test the performance against a single loss anomaly value or its
position, we varied them in the previous three-span system. Figure
\ref{fig:var_lossan_zan} depicts the results in terms of the RMS
error, which is still restricted to losses smaller than $15$ dB.
We observe an excellent behavior with respect to both variables, with
minor impact of the SNR in the selected ranges.

\begin{figure}[tbh]
\begin{centering}
\includegraphics[width=0.8\linewidth]{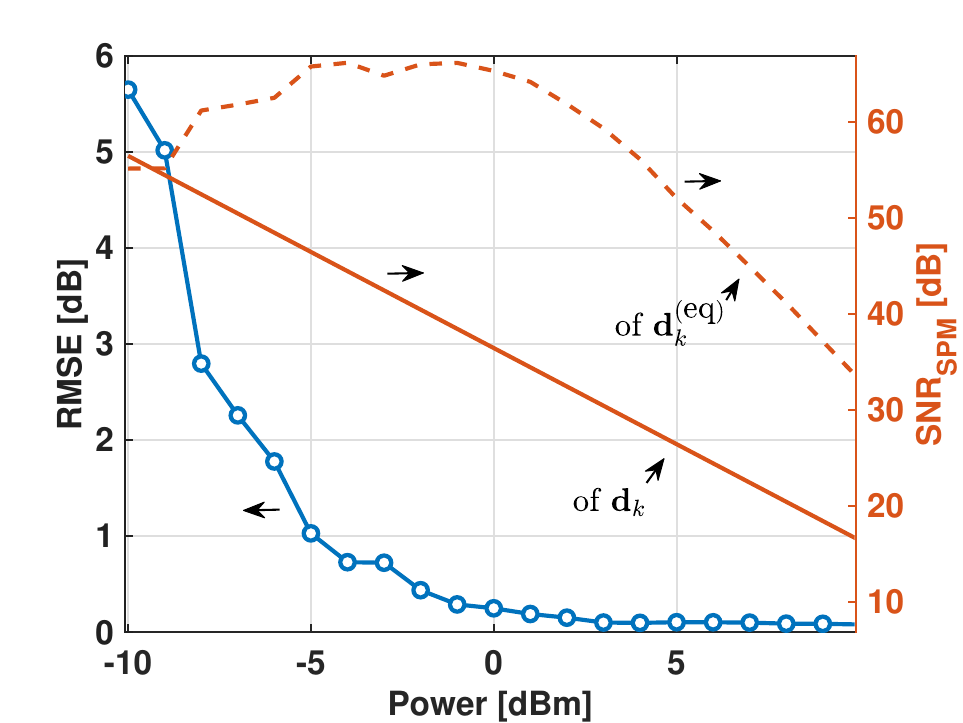}
\par\end{centering}
\caption{(Left) RMSE vs. signal power with a residual noise SNR of $20$ dB.
(Right) SNR associated with SPM only experienced by the detected symbols
$\mathbf{d}_{k}$ or equalized symbols $\mathbf{d}_{k}^{(\text{eq})}$.
The results are for a three-span link\label{fig:varpower}}

\end{figure}

Next we tested the impact of the signal power. Figure \ref{fig:varpower}
illustrates the algorithm behavior, both in terms of the RMSE and
SPM compensation capabilities. Here we used SNR=$20$ dB. The primary
performance metric, the RMSE, is shown on the left y-axis. We observe
a failure of the method at very low power, where the residual noise
dominates the NLI power due to SPM. However, it is worth noting that
the algorithm is capable of detecting the useful signal, SPM, even
if the SNR associated to the residual noise is smaller than the SNR
associated to SPM, as visible in the right-axis. We also reported
for comparison the SNR of SPM after nonlinear equalization, hence
by computing the noise power on the residual NLI of $\mathbf{d}_{k}^{(\text{eq})}$.
We observe a significant reduction of SPM with the chosen grid of
$5$ km, highlighting the potential for SPM compensation performed
together with monitoring.

\begin{figure}[tbh]
\begin{centering}
\includegraphics[width=0.8\linewidth]{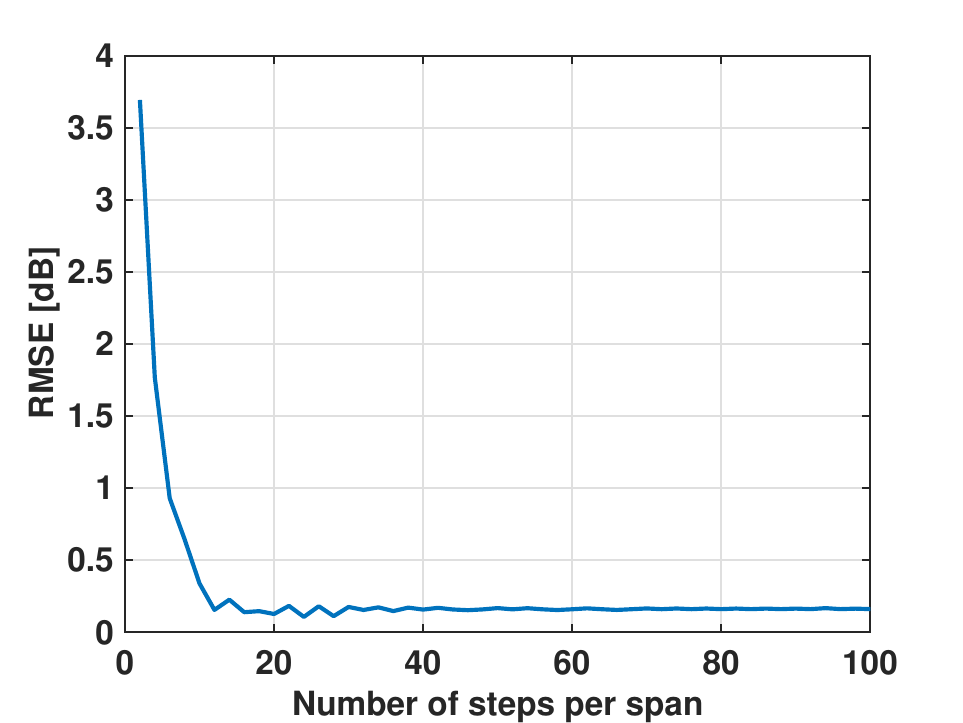}
\par\end{centering}
\caption{RMSE vs the number of steps per span adopted in the digital twin.
The results are for a three-span link with a residual noise SNR of
$10$ dB.\label{fig:vargrid}}

\end{figure}

In another test we investigated the impact of the spatial grid used
in the digital twin, see (\ref{eq:nliquad}). We estimated the RMS
error vs the number of steps per span in the three-span system, here
at SNR=$10$ dB. The result is shown in Fig. \ref{fig:vargrid}, which
we remind is based on a mid-point numerical integration rule. The
anomaly was at $125$ km. Clearly, we see a decline in performance
with fewer steps per span. The explanation is related to the the digital
twin's inability to reproduce the real system. Therefore, if computational
complexity is a concern, it may be beneficial to explore more efficient
quadrature rules for numerical integration. As a rough estimate, we
observe a severe degradation of performance around $\approx14$ steps
per span (step size $\approx7$ km), corresponding to a phase shift
of GVD in the frequency domain of $\pi$ at the signal 3dB bandwidth.

\begin{figure}[tbh]

\begin{centering}
\includegraphics[width=0.9\linewidth]{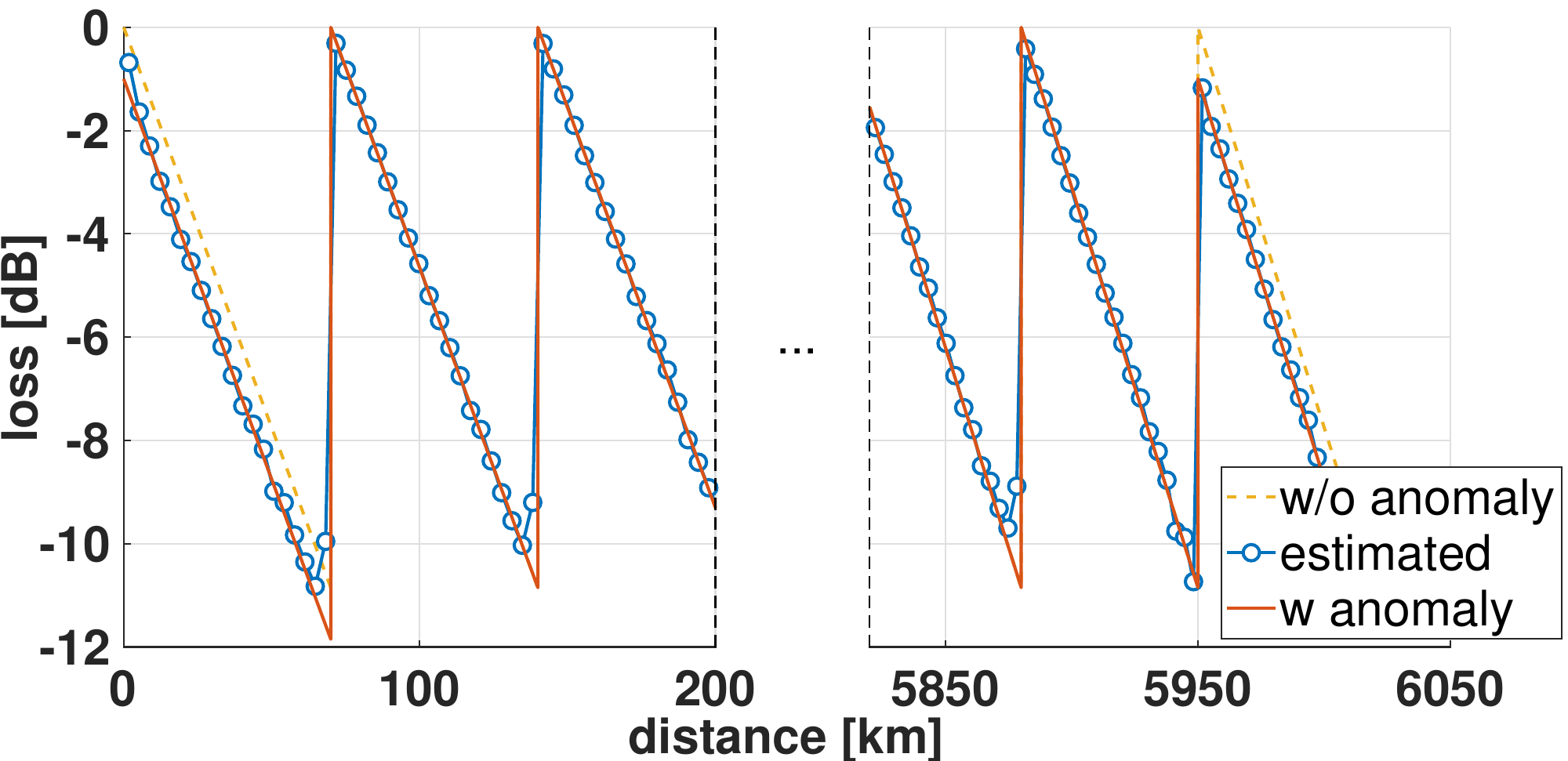}
\par\end{centering}
\caption{Subsea link of $6020$ km with two loss anomalies at the beginning
of the first and the last span, respectively. SNR=$10$ dB.\label{fig:Subsea-link-of}}

\end{figure}

Having tested the algorithm in terrestrial links, we then moved to
a subsea link. We examined a transatlantic $86$-span with $70$ km
per span (total $6020$ km), with two loss anomalies of $1$ dB at
zero and $5950$ km, i.e., at the beginning of the first and the last
span. We changed the link parameters to suit the different scenario,
hence we used dispersion $21$ $\text{ps}\cdot\text{nm}^{-1}\text{km}^{-1}$,
nominal loss $0.155$ dB/km, $\gamma=0.92$ W$^{-1}$km$^{-1}$, and
channel power $-1$ dBm. The SNR was $10$ dB, a realistic number
for such a system. Figure \ref{fig:Subsea-link-of} shows the loss
profile estimated with the LMS after $32$M symbols and the true one,
observing an excellent match. It is worth noting that in this ultralong
case we had to reduce $\bar{\mu}$ to $0.01$. In the subsea case,
the computational effort is significant because both the GVD memory,
thus the block size $L$, and the spatial grid size grow linearly
with distance.

\begin{figure}[tbh]
\begin{centering}
\includegraphics[width=0.8\linewidth]{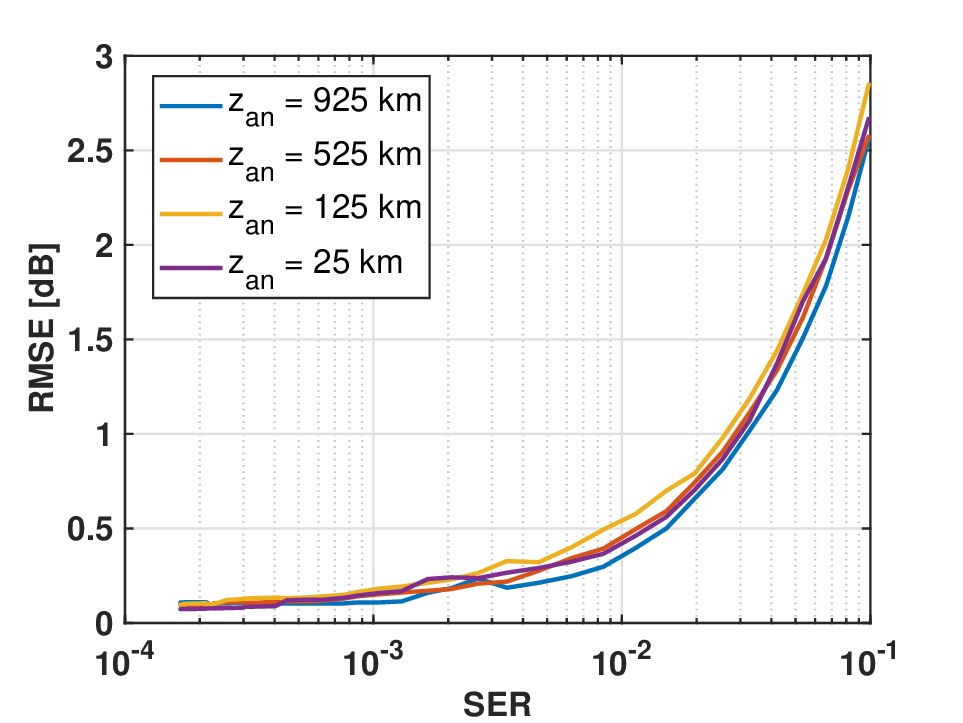}
\par\end{centering}
\caption{RMSE vs symbol error rate. Decision-directed LMS without FEC. Loss
anomaly of $1$ dB at different positions ($z_{\text{an}}$) of a
$10\times100$ km link.\label{fig:RMS-vs-SER}}
\end{figure}

The previous results were in data-aided mode. To check the impact
of wrong decisions we tested the behavior by feeding the digital twin
with the pre-FEC decided data. As a quality factor, we evaluated the
RMSE versus symbol error rate (SER), as shown in Fig. \ref{fig:RMS-vs-SER}.
The SER was controlled by varying the residual noise SNR. Here we
analyzed a $10\times100$ km link with a loss anomaly placed at various
positions, $z_{\text{an}}$ in the figure. The results show that algorithm
performance worsens as SER increases, regardless of the position of
the anomaly, with a maximum tolerable SER of around $10^{-3}$. We
thus deduce that longer connections suffer more from this requirement.

\section{Conclusions\label{sec:Conclusions}}

We proposed a block LMS technique for monitoring the longitudinal
power profile of a fiber-optic link. The proposed technique is efficient,
based on a full digital twin of the link, and is adaptive by nature,
using a stochastic gradient descent updating rule based on a technique
in the frequency domain. We process data by using an overlap-save
algorithm, avoiding the need for batch processing of large datasets.
The proposed method leverages the main advantages of the LMS algorithm
in the context of linear adaptive filter theory, like its simplicity
and small memory footprint. Compared with the least squares method,
there is no need to invert a large matrix. The proposed method is
able to identify multiple loss anomalies without the need of a calibration
stage. 

In its current version, the algorithm just requires knowledge of fiber
dispersion. Future research will focus on expanding the digital twin
to include more complex optical devices.

\appendices{}

\section{Overlap-save Algorithm\label{sec:Overlap-save-Algorithm}}

In this appendix we review the basic properties of the overlap-save
algorithm \cite{Manolakis,Haykin} and specialize them to our framework.
Let $y(i)$ be the output of the linear discrete convolution $\otimes$
between a FIR filter of $L$ samples and a longer signal $x$:
\[
y(i)=(h\otimes x)(i)=\sum_{n}h(n)x(i-n)\,.
\]
The overlap-save algorithm is an efficient method to compute $y(kL+i)$
for a given integer $k$ and $i=0,\ldots,L-1$, i.e., to compute $y$
within the $k$th block of $L$ samples. The key relationship at the
hearth of the method is:
\begin{equation}
\left(h\otimes x\right)(kL+i)\equiv\left(h_{\text{pad}}\circledast x_{k}\right)_{\text{val}}\left(i\right),\quad i=0,\ldots,L-1\label{eq:key_os}
\end{equation}
where $\circledast$ indicates circular convolution, $x_{k}$ is a
vector of $2L$ samples resulting from the concatenation of the input
signal in blocks $(k-1,k)$:
\begin{equation}
x_{k}\triangleq x(i),\quad i=(k-1)L,\ldots,(k+1)L-1\,.\label{eq:xck}
\end{equation}
The subscript ``val'' denotes the valid, not-aliased part of the
circular convolution result, which corresponds to the latter half
of the output vector. However, keep in mind that a processing delay
is inherent when, like in our case, the digital twin models the real
system using a FIR implementation of non-causal filters. $h_{\text{pad}}$
is:
\begin{align}
h_{\text{pad}}(i) & \triangleq\begin{cases}
h(i) & i=0,1,\ldots,L-1\\
0 & i=L,\ldots,2L-1\,.
\end{cases}\label{eq:hpad}
\end{align}
Note that it is not necessary to split $x$ into blocks of $L$ samples.
Longer blocks imply redundant operations, whereas shorter blocks reduce
processing latency \cite{Haykin}. Setting the block length equal
to $L$ is generally considered as an optimal choice.

\end{document}